\documentclass[12pt]{article}
\usepackage{epsfig}

\def\beq{\begin{equation}}
\def\eeq#1{\label{#1}\end{equation}}
\def\eeqn{\end{equation}}

\def\beqa{\begin{eqnarray}}
\def\eeqa#1{\label{#1}\end{eqnarray}}
\def\eeqan{\end{eqnarray}}

\let\bar=\overbar

\def\D{{\cal D}}

\def\Dslash{\not{\hbox{\kern-4pt $D$}}}
\def\dslash{\not{\hbox{\kern-2pt $\del$}}}

\def\msb{{\bar{\ssstyle M \kern -1pt S}}}

\def\Title#1{\begin{center} {\Large {\bf #1} } \end{center}}

\usepackage{xcolor}
\usepackage{ifthen} 
\usepackage{booktabs}

\newboolean{pdflatex}
\setboolean{pdflatex}{true} 

\newboolean{articletitles}
\setboolean{articletitles}{true} 

\newboolean{uprightparticles}
\setboolean{uprightparticles}{false} 

\newboolean{inbibliography}
\setboolean{inbibliography}{false} 

\usepackage[top=1in, bottom=1.25in, left=1in, right=1in]{geometry}

\usepackage{microtype}
\usepackage{lineno}  
\usepackage{xspace} 
\usepackage{caption} 

\usepackage{graphicx}  
\usepackage{color}
\usepackage{colortbl}
\graphicspath{{./figs/}} 

\usepackage{amsmath} 
\usepackage{amssymb}
\usepackage{amsfonts}
\usepackage{upgreek} 

\newcommand*\patchAmsMathEnvironmentForLineno[1]{%
\expandafter\let\csname old#1\expandafter\endcsname\csname #1\endcsname
\expandafter\let\csname oldend#1\expandafter\endcsname\csname
end#1\endcsname
 \renewenvironment{#1}%
   {\linenomath\csname old#1\endcsname}%
   {\csname oldend#1\endcsname\endlinenomath}%
}
\newcommand*\patchBothAmsMathEnvironmentsForLineno[1]{%
  \patchAmsMathEnvironmentForLineno{#1}%
  \patchAmsMathEnvironmentForLineno{#1*}%
}
\AtBeginDocument{%
\patchBothAmsMathEnvironmentsForLineno{equation}%
\patchBothAmsMathEnvironmentsForLineno{align}%
\patchBothAmsMathEnvironmentsForLineno{flalign}%
\patchBothAmsMathEnvironmentsForLineno{alignat}%
\patchBothAmsMathEnvironmentsForLineno{gather}%
\patchBothAmsMathEnvironmentsForLineno{multline}%
\patchBothAmsMathEnvironmentsForLineno{eqnarray}%
}

\usepackage{hyperref}    
\usepackage[all]{hypcap} 

\usepackage{xspace} 
\usepackage{upgreek}

\def\lhcb {\mbox{LHCb}\xspace}

\def\MagUp {\mbox{\em Mag\kern -0.05em Up}\xspace}

\ifthenelse{\boolean{uprightparticles}}%
{

 \def\Ppi         {\ensuremath{\uppi}\xspace}

 \def\PDelta      {\ensuremath{\Delta}\xspace}                 
 \def\PXi      {\ensuremath{\Xi}\xspace}                 
 \def\PLambda      {\ensuremath{\Lambda}\xspace}                 
 \def\PSigma      {\ensuremath{\Sigma}\xspace}                 
 \def\POmega      {\ensuremath{\Omega}\xspace}                 
 \def\PUpsilon      {\ensuremath{\Upsilon}\xspace}                 
 
 \def\PB      {\ensuremath{\mathrm{B}}\xspace}                 
                  
 \def\PD      {\ensuremath{\mathrm{D}}\xspace}

 \def\PK      {\ensuremath{\mathrm{K}}\xspace}

 \def\Pb      {\ensuremath{\mathrm{b}}\xspace}                 
 \def\Pc      {\ensuremath{\mathrm{c}}\xspace}

 \def\Pi      {\ensuremath{\mathrm{i}}\xspace}

 \def\Ps      {\ensuremath{\mathrm{s}}\xspace}

}
{

 \def\Ppi         {\ensuremath{\pi}\xspace}

 \mathchardef\PDelta="7101
 \mathchardef\PXi="7104
 \mathchardef\PLambda="7103
 \mathchardef\PSigma="7106
 \mathchardef\POmega="710A
 \mathchardef\PUpsilon="7107
                  
 \def\PB      {\ensuremath{B}\xspace}                 
                  
 \def\PD      {\ensuremath{D}\xspace}

 \def\PK      {\ensuremath{K}\xspace}

 \def\Pb      {\ensuremath{b}\xspace}                 
 \def\Pc      {\ensuremath{c}\xspace}

 \def\Pi      {\ensuremath{i}\xspace}

 \def\Ps      {\ensuremath{s}\xspace}

}

\makeatletter
\ifcase \@ptsize \relax
  \newcommand{\miniscule}{\@setfontsize\miniscule{4}{5}}
\or
  \newcommand{\miniscule}{\@setfontsize\miniscule{5}{6}}
\or
  \newcommand{\miniscule}{\@setfontsize\miniscule{5}{6}}
\fi
\makeatother

\DeclareRobustCommand{\optbar}[1]{\shortstack{{\miniscule (\rule[.5ex]{1.25em}{.18mm})}
  \\ [-.7ex] $#1$}}

\def\squark    {{\ensuremath{\Ps}}\xspace}

\def\cquark    {{\ensuremath{\Pc}}\xspace}

\def\bquark    {{\ensuremath{\Pb}}\xspace}

\def\pion   {{\ensuremath{\Ppi}}\xspace}

\def\pim    {{\ensuremath{\pion^-}}\xspace}

\def\kaon    {{\ensuremath{\PK}}\xspace}
  \def\Kbar    {{\kern 0.2em\overline{\kern -0.2em \PK}{}}\xspace}

\def\KorKbar    {\kern 0.18em\optbar{\kern -0.18em K}{}\xspace}
\def\Kz      {{\ensuremath{\kaon^0}}\xspace}

  \def\Dbar    {{\kern 0.2em\overline{\kern -0.2em \PD}{}}\xspace}
\def\D       {{\ensuremath{\PD}}\xspace}

\def\DorDbar    {\kern 0.18em\optbar{\kern -0.18em D}{}\xspace}
\def\Dz      {{\ensuremath{\D^0}}\xspace}

\def\Dm      {{\ensuremath{\D^-}}\xspace}

\def\Dstarm  {{\ensuremath{\D^{*-}}}\xspace}

\def\Dsp     {{\ensuremath{\D^+_\squark}}\xspace}
\def\Dsm     {{\ensuremath{\D^-_\squark}}\xspace}

\def\Dssm    {{\ensuremath{\D^{*-}_\squark}}\xspace}

\def\B       {{\ensuremath{\PB}}\xspace}
\def\Bbar    {{\ensuremath{\kern 0.18em\overline{\kern -0.18em \PB}{}}}\xspace}

\def\BorBbar    {\kern 0.18em\optbar{\kern -0.18em B}{}\xspace}
\def\Bz      {{\ensuremath{\B^0}}\xspace}

\def\Bu      {{\ensuremath{\B^+}}\xspace}

\def\Bp      {{\ensuremath{\Bu}}\xspace}

\def\Bs      {{\ensuremath{\B^0_\squark}}\xspace}
\def\Bsb     {{\ensuremath{\Bbar{}^0_\squark}}\xspace}

\def\Bc      {{\ensuremath{\B_\cquark^+}}\xspace}

  \def\Y#1S{\ensuremath{\PUpsilon{(#1S)}}\xspace}

\def\Lz          {{\ensuremath{\PLambda}}\xspace}
\def\Lbar        {{\ensuremath{\kern 0.1em\overline{\kern -0.1em\PLambda}}}\xspace}
\def\LorLbar    {\kern 0.18em\optbar{\kern -0.18em \PLambda}{}\xspace}

\def\Lb      {{\ensuremath{\Lz^0_\bquark}}\xspace}

\def\Lc      {{\ensuremath{\Lz^+_\cquark}}\xspace}

\def\to                 {\ensuremath{\rightarrow}\xspace}

\def\AT#1     {\ensuremath{A_{\mathrm{T}}^{#1}}\xspace}           

\def\C#1      {\ensuremath{\mathcal{C}_{#1}}\xspace}                       
\def\Cp#1     {\ensuremath{\mathcal{C}_{#1}^{'}}\xspace}                    
\def\Ceff#1   {\ensuremath{\mathcal{C}_{#1}^{\mathrm{(eff)}}}\xspace}        
\def\Cpeff#1  {\ensuremath{\mathcal{C}_{#1}^{'\mathrm{(eff)}}}\xspace}       
\def\Ope#1    {\ensuremath{\mathcal{O}_{#1}}\xspace}                       
\def\Opep#1   {\ensuremath{\mathcal{O}_{#1}^{'}}\xspace}                    

\newcommand{\tev}{\ifthenelse{\boolean{inbibliography}}{\ensuremath{~T\kern -0.05em eV}}{\ensuremath{\mathrm{\,Te\kern -0.1em V}}}\xspace}
\newcommand{\gev}{\ensuremath{\mathrm{\,Ge\kern -0.1em V}}\xspace}
\newcommand{\mev}{\ensuremath{\mathrm{\,Me\kern -0.1em V}}\xspace}
\newcommand{\kev}{\ensuremath{\mathrm{\,ke\kern -0.1em V}}\xspace}
\newcommand{\ev}{\ensuremath{\mathrm{\,e\kern -0.1em V}}\xspace}
\newcommand{\gevc}{\ensuremath{{\mathrm{\,Ge\kern -0.1em V\!/}c}}\xspace}
\newcommand{\mevc}{\ensuremath{{\mathrm{\,Me\kern -0.1em V\!/}c}}\xspace}
\newcommand{\gevcc}{\ensuremath{{\mathrm{\,Ge\kern -0.1em V\!/}c^2}}\xspace}
\newcommand{\gevgevcccc}{\ensuremath{{\mathrm{\,Ge\kern -0.1em V^2\!/}c^4}}\xspace}
\newcommand{\mevcc}{\ensuremath{{\mathrm{\,Me\kern -0.1em V\!/}c^2}}\xspace}

\def\mum  {\ensuremath{{\,\upmu\mathrm{m}}}\xspace}

\def\invfb   {\ensuremath{\mbox{\,fb}^{-1}}\xspace}

\def\invps{\ensuremath{{\mathrm{ \,ps^{-1}}}}\xspace}

\newcommand{\stat}{\ensuremath{\mathrm{\,(stat)}}\xspace}
\newcommand{\syst}{\ensuremath{\mathrm{\,(syst)}}\xspace}

\def\gsim{{~\raise.15em\hbox{$>$}\kern-.85em
          \lower.35em\hbox{$\sim$}~}\xspace}
\def\lsim{{~\raise.15em\hbox{$<$}\kern-.85em
          \lower.35em\hbox{$\sim$}~}\xspace}

\def\pt         {\mbox{$p_{\mathrm{ T}}$}\xspace}

\def\tell1  {TELL1\xspace}
\def\ukl1   {UKL1\xspace}

\usepackage{cite} 
\usepackage{mciteplus}

\usepackage{longtable} 

\usepackage{overpic}
\usepackage{etoolbox}

\def\DSSm    {{\ensuremath{\D^{\scalebox{0.4}{(}*\scalebox{0.4}{)}-}_{\squark}}}\xspace}
\def\DSm    {{\ensuremath{\D^{\scalebox{0.4}{(}*\scalebox{0.4}{)}-}}}\xspace}
\def\DSSM    {{\ensuremath{\D^{\scalebox{0.4}{(}*\scalebox{0.4}{)}-}_{(\squark)}}}\xspace}
\def\DSM    {{\ensuremath{\D^{\scalebox{0.4}{}\scalebox{0.4}{}-}_{(\squark)}}}\xspace}
\def\DSP    {{\ensuremath{\D^{\scalebox{0.4}{}\scalebox{0.4}{}+}_{(\squark)}}}\xspace}

\robustify{\DSm}
\robustify{\DSSm}
\robustify{\DSSM}
\robustify{\DSM}
\robustify{\DSP}

\newcommand{\taufs}{\ensuremath{\tau^{\rm fs}_{\Bs}}\xspace}

\newcommand{\dB}{\ensuremath{\rm -0.0115} }
\newcommand{\eStatdB}{\ensuremath{0.0053}}

\newcommand{\eSystdB}{\ensuremath{0.0041}}

\newcommand{\tB}{\ensuremath{\rm 1.547} }
\newcommand{\eStattB}{\ensuremath{0.013} } 
\newcommand{\eSysttB}{\ensuremath{0.010} } 
\newcommand{\eReftB}{\ensuremath{0.004} }
\newcommand{\refB}{\ensuremath{\,(\tau_{\it B})}} 

\newcommand{\dD}{\ensuremath{1.0131} }
\newcommand{\eStatdD}{\ensuremath{0.0117} }

\newcommand{\eSystdD}{\ensuremath{0.0065} }
 
\newcommand{\tD}{\ensuremath{0.5064} } 
\newcommand{\eStattD}{\ensuremath{0.0030} }

\newcommand{\eSysttD}{\ensuremath{0.0017} }
\newcommand{\eReftD}{\ensuremath{0.0017} }
\newcommand{\refD}{\ensuremath{\,(\tau_{\it D})}}

\begin{document}

\begin{raggedright}  

{Talk presented at the APS Division of Particles and Fields Meeting (DPF 2017), July 31-August 4, 2017, Fermilab. C170731}
\bigskip\bigskip
\end{raggedright}

\Title{A novel measurement of $B^0_s$ and $D^-_s$ lifetimes using\\
\vspace{.2cm} semileptonic decays at LHCb}

\bigskip\bigskip

\begin{raggedright}  

{\it Diego Tonelli\index{Tonelli, D.} for the LHCb Collaboration\\
INFN Sezione di Trieste\\
Padriciano, 99\\
I-34149 Trieste, Italy}
\bigskip\bigskip
\end{raggedright}

\begin{abstract}
I report new, world-leading  LHCb results on heavy meson lifetimes. We use a novel approach that suppresses the shortcomings typically associated with reconstruction of semileptonic decays, allowing for precise measurements of lifetimes and other properties in collider experiments. We achieve a 15\% and a $2\times$ improvement over the current best determinations of the flavor-specific \Bs lifetime and \Dsm lifetime, respectively.

\end{abstract}

\section{Heavy hadron lifetimes}

Lifetimes are fundamental properties of particles, which connect deeply with their dynamics. Improved lifetime determinations of heavy hadrons probe the interplay of the strong and weak interactions between constituent partons, stimulating further refinement of the phenomenological understanding.  Most importantly, measurements of heavy hadron lifetimes enhance the reach in indirect searches for non-standard-model physics. Comparisons of similarly precise measurements and predictions of observables associated with quark-flavor dynamics probe the existence of non-standard-model particles of masses much larger than those directly accessible at particle colliders. The precision of the predictions is often limited by difficulties in calculating strong-interaction transition amplitudes at low energies. Predictability is often recovered by resorting to effective models such as heavy-quark expansion~\cite{Lenz:2014jha}. Heavy-hadron lifetimes offer precious and constraining validation and tuning of such models. \par Precise \Bs\ lifetime measurements are particularly needed. In fact, the \Bs\ lifetime precision has a significant impact in the  lifetime ratio between \Bs and \Bz mesons, which shows a 2.5 standard-deviation discrepancy from predictions that calls for further investigation. Especially relevant are measurements of the ``flavor-specific" \Bs meson lifetime, 
\begin{equation}
\taufs\equiv\frac{1}{\Gamma_s} \left[ \frac{1+(\Delta\Gamma_s/2\Gamma_s)^2}{1-(\Delta\Gamma_s/2\Gamma_s)^2} \right],
\end{equation}
where $\Gamma_s = (\Gamma_{s,H} + \Gamma_{s,L})/2$ and $\Delta\Gamma_s = \Gamma_{s,L} - \Gamma_{s,H}$ are the average and the difference, respectively, of the natural widths $\Gamma_{s,H(L)}$ of the heavy (light) mass eigenstate. This empirical quantity allows an indirect determination of $\Delta\Gamma_s$ that, compared with direct determinations, may test the presence of non-standard-model physics~\cite{HFAG}. The lifetime $\taufs$ is measured with a single-exponential fit to the distribution of decay time to a final state not accessible by both \Bs and \Bsb mesons~\cite{Hartkorn:1999ga}. The current best determination, $\taufs = 1.535 \pm 0.015 ({\rm stat}) \pm 0.014({\rm syst}) $\,ps~\cite{LHCb-PAPER-2014-037}, obtained by the LHCb collaboration using hadronic $\Bs \to \Dsm \pi^+$ decays, has similarly-sized statistical and systematic uncertainties. Throughout this document, the symbol $X$ identifies any decay product, other than neutrinos, not included in the candidate reconstruction, and the inclusion of charge-conjugate processes is implied. \par Semileptonic \Bs decays, owing to larger signal yields than from hadronic decays, offer richer potential for precise $\taufs$ measurements. However, neutrinos and other low-momentum neutral final-state particles prevent the full reconstruction of such decays. This introduces serious limitations due to degraded understanding of background contributions and difficulties in obtaining the decay time from the observed decay-length distribution. Measurements of bottom-meson lifetimes using semileptonic decays, which had been popular at LEP, $B$-factories, and Tevatron Run I since the '90s through approximately 2004--2006, became rarer afterwards, when large samples of fully reconstructed $B \to J/\psi X$ decays become available. Controlling systematic uncertainties proved challenging~\cite{daveclark,satoru} and rarely  analyses achieved competitive results, which were anyhow limited by the size of the systematic uncertainty~\cite{Bc,Abazov:2014rua}. 
\par The LHCb Collaboration has recently proposed a novel, data-driven approach that suppresses such limitations thus achieving a world-class measurement of \taufs  with small systematic uncertainty~\cite{paper}.  The analysis also yields a strongly improved determination of the \Dsm lifetime over the current best result, $\tau_{\Dsm}= 0.5074 \pm 0.0055\stat \pm 0.0051\syst$ ps, reported more than a decade ago by the FOCUS collaboration~\cite{Link:2005ew}. Such a novel analysis approach is not necessarily restricted to LHCb or to determinations of lifetimes solely. 
\section{Overview}
The \Bs and $D^-_s$ lifetimes are determined from the variation in \Bs signal yield as a function of decay time, relative to that of \Bz decays reconstructed in the same final state. The use of kinematically similar \Bz decays of precisely known lifetime, as a reference, suppresses the uncertainties from partial reconstruction and lifetime-biasing selection criteria. \par 
We analyze proton-proton collisions at center-of-mass energies of 7 and 8 TeV collected by the LHCb experiment in 2011 and 2012 and corresponding to an integrated luminosity of 3.0 fb$^{-1}$. We reconstruct approximately 407\,000 $\Bs \to \Dssm \mu^+\nu_\mu$ and $\Bs \to \Dsm \mu^+\nu_\mu$ ``signal" decays, and approximately 108\,000 $\Bz \to \Dstarm \mu^+\nu_\mu$ and $\Bz \to \Dm \mu^+\nu_\mu$ ``reference" decays.  The $D$ candidates are reconstructed as combinations of $K^+$, $K^-$, and $\pi^-$ candidates originating from a common space-point (vertex), displaced from any proton-proton interaction vertex. The $B^0_{(s)}$ candidates, namely $K^+ K^- \pi^- \mu^+$ combinations,  are formed by $D$ candidates associated with muon candidates originating from another common displaced vertex.  We collectively refer to the signal and reference decays as $\Bs \to [K^+K^-\pi^-]_\DSSm \mu^+\nu_\mu$ and $\Bz \to [K^+K^-\pi^-]_\DSm \mu^+\nu_\mu$, respectively. A fit to the ratio of event yields between the signal and reference  decays as a function of $B^0_{(\squark)}$ decay time, $t$,  determine $\Delta_\Gamma(B) \equiv 1/\taufs - \Gamma_d$, where $\Gamma_d$ is the known natural width of the \Bz meson. A similar fit, performed as a function of the $D^-_{(s)}$ decay time, determines the decay-width difference between \Dsm and \Dm mesons, $\Delta_\Gamma(D)$. Event yields are determined by fitting the candidates' ``corrected-mass" distribution, $m_{\rm corr} = p_{\perp, D\mu} + \sqrt{m^2_{D\mu}+p^2_{\perp, D\mu}}$~\cite{Kodama:1991ij}. The corrected mass is determined from the invariant mass of the $D_{(s)}^-\mu^+$ pair, $m_{D\mu}$, and the component of its momentum perpendicular to the $B^0_{(s)}$ flight direction, $p_{\perp, D\mu}$, to compensate for the average transverse momentum of unreconstructed decay products. The flight direction is the directed line-segment connecting the $B^0_{(s)}$ production and decay vertices; the decay time $t = m_B L k/ p_{D\mu}$ is calculated from the known $B^0_{(s)}$ mass, $m_B$~\cite{PDG2016}, the observed $B^0_{(s)}$ decay length, $L$,  and the $D^-_{(s)}\mu^+$-pair momentum, $p_{D\mu}$. The scale factor $k$ corrects $p_{D\mu}$ for the average momentum fraction carried by decay products excluded from the reconstruction~\cite{Abulencia:2006ze, Leonardo:2006fq}. Decay-time acceptances and resolutions, determined from simulation, are included in the fits.
\section{LHCb detector and simulation}
The LHCb detector~\cite{Alves:2008zz,Aaij:2014jba} is a single-arm forward spectrometer covering $2 < \eta < 5$ pseudorapidity, designed for the study of particles containing bottom or charm quarks. The detector allows tracking using a silicon-strip vertex detector surrounding the interaction region, a large-area silicon-strip detector located upstream of a dipole magnet with a bending power of about 4 Tm, and three stations of silicon-strip detectors and straw drift tubes installed downstream of the magnet. The fractional resolution on charged-particle's momentum $p$ is 0.5\%--1.0\%. The minimum distance of a charged-particle trajectory (track) to a primary vertex, the impact parameter, is measured with $(15 + 29/p_T)$ $\mum$ resolution,  where $p_T$ is the $p$ component transverse to the beam, in GeV/$c$. Charged-hadron species are distinguished using two ring-imaging Cherenkov detectors. Photons, electrons and hadrons are identified by a sampling calorimeter consisting of scintillating-pad electromagnetic and hadronic portions and preshower detectors. Muons are identified using alternating layers of iron and multiwire proportional chambers. The online event selection is performed by a hardware trigger, based on information from the calorimeter and muon systems, followed by a software trigger, which applies a full event reconstruction.  Simulation of collisions is provided by a specially configured {\sc Pythia} software package. Hadron decays are described by {\sc EvtGen} including final-state radiation simulated using {\sc Photos}. The interaction of particles with the detector and its response are simulated using the {\sc geant4} toolkit~\cite{LHCb-PROC-2011-006, LHCb-PROC-2010-056}.  Simulation is used to identify all relevant sources of bottom-hadron decays, model the mass distributions, and correct for the effects of incomplete kinematic reconstructions, relative decay-time acceptances, and decay-time resolutions. The unknown details of the \Bs decay dynamics are modeled in the simulation through empirical form-factor parameters~\cite{Caprini:1997mu}, assuming values inspired by the known $B^0$ form factors~\cite{HFAG}. The impact of these assumptions is accounted for in the systematic uncertainties.  
\section{Sample selection}
The trigger requires a muon candidate, with $\pt > 1.5-1.8$\gevc, associated with 1--3 charged particles, all originating in a vertex displaced from the proton-proton vertex~\cite{Aaij:2012me} and pointing to the displaced vertex where the muon candidate originates from. \par
\begin{figure}[t]
\centering
\includegraphics[width=0.48\textwidth]{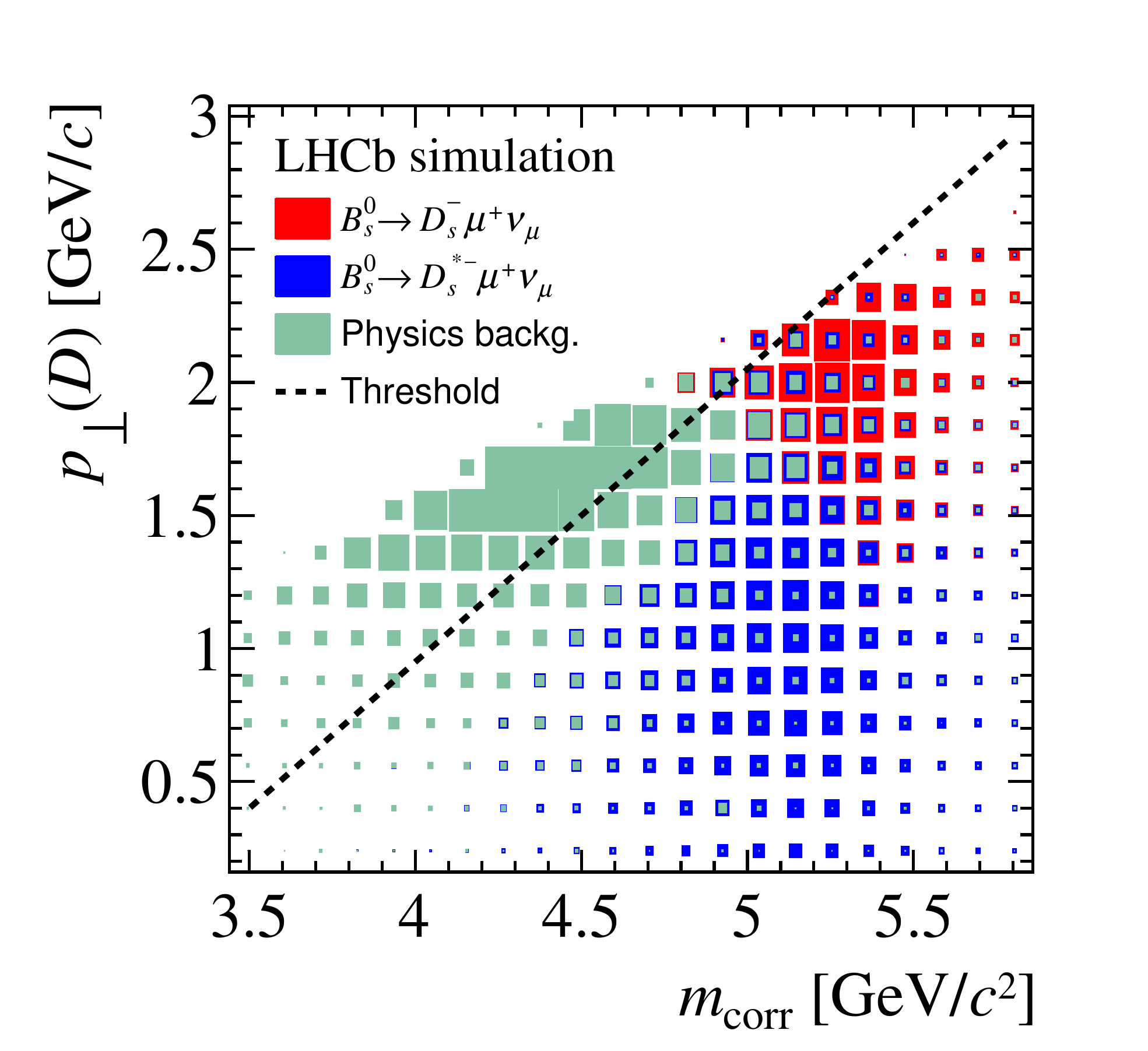}\\
\caption{Two-dimensional distribution of the $\D^-_{(s)}$ momentum-component perpendicular to the $\B^0_{(\squark)}$ flight direction as a function of $m_{\rm corr}$ for three classes of simulated events. The linear boundary used in the analysis is represented by the dashed line.\label{fig:threshold}}
\end{figure}

 Offline, the muon is combined with charged particles consistent with the topology and kinematics of signal $\Bs \to [K^+K^-\pi^-]_\DSSm \mu^+\nu_\mu$ and reference  $\Bz \to [K^+K^-\pi^-]_\DSm \mu^+\nu_\mu$ decays.  The accepted $K^+K^-\pi^-$ mass range is restricted around the known \DSM meson masses to suppress signal-reference cross-contamination to less than 0.1\%, as estimated from simulation.  We also reconstruct ``same-sign" $K^+K^-\pi^-\mu^-$ candidates, formed by charm and muon candidates with same-sign charge, to model combinatorial background from accidental $D^{-}_{(s)} \mu^+$ associations. The event selection is designed to suppress the background under the charm signals and making same-sign candidates a reliable model for the combinatorial background: track- and vertex-quality, vertex-displacement, \pt, and particle-identification criteria are chosen such as to minimize shape and yield differences between same-sign and signal candidates in the $m_{D\mu} > 5.5\gevcc$ region, where genuine bottom-hadron decays are kinematically excluded and combinatorial background dominates. Mass vetoes suppress background from misreconstructed decays such as $\Bs \to \psi^{(')}(\to \mu^+\mu^-)\phi (\to K^+K^-)$ decays where a muon is misidentified as a pion, $\Lb \to \Lc (\to pK^-\pi^+) \mu^- \bar{\nu}_\mu X$ decays where the proton is misidentified as a kaon or a pion, and $B^0_{(s)} \to D^-_{(s)}\pi^+$ decays where the pion is misidentified as a muon. Significant contributions arise from decays of a bottom hadron into pairs of charm hadrons, one peaking at the $D^-_{(s)}$ mass and the other decaying semileptonically, or into single charm hadrons and other particles.  Such decays include $\B^0_{(\squark)} \to 
\D^{\scalebox{0.4}{(}*\scalebox{0.4}{)}-}_{(\squark)}\DSP$,  $\Bp \to \Dbar{}^{\scalebox{0.4}{(}*\scalebox{0.4}{)}0} D^{\scalebox{0.4}{(}*\scalebox{0.4}{)}+}$, $\Bp \to \Dm \mu^+ \nu_\mu X$, $\Bp \to \DSSm K^+ \mu^+ \nu_\mu X$, $\Bz \to \DSSm \Kz \mu^+ \nu_\mu X$, $\Bs \to \Dz\Dsm K^+$, $\Bs \to \Dm \Dsp \Kz$, $\Lb \to \Lc \DSSm X$, and $\Lb \to \Dsp \Lz \mu^- \bar{\nu}_\mu X$ decays. We suppress these backgrounds with an upper threshold, linearly dependent on $m_{\rm corr}$, applied to the $\D^-_{(s)}$ momentum component perpendicular to the $\B^0_{(\squark)}$ flight direction, shown in Fig.~\ref{fig:threshold}.  Finally, a $t>0.1$\,ps requirement on the $\D^-_{(\squark)}$ proper decay time renders the signal- and reference-decay acceptances as functions of decay time more similar, with little signal loss.
\section{Data analysis}
\begin{figure}
\centering
\begin{overpic}[width=0.48\textwidth]{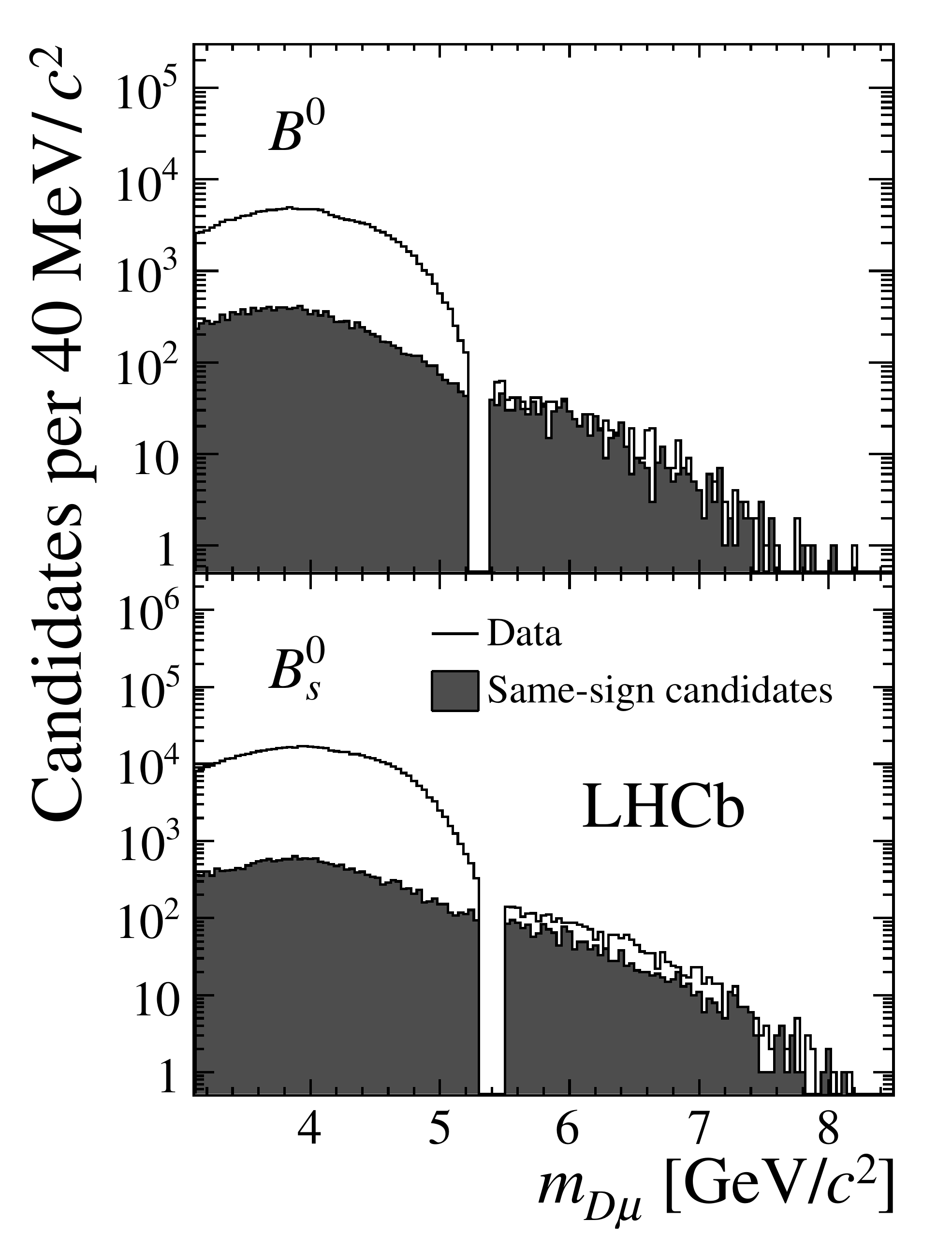}
\put(41,69){\includegraphics[width=0.17\textwidth]{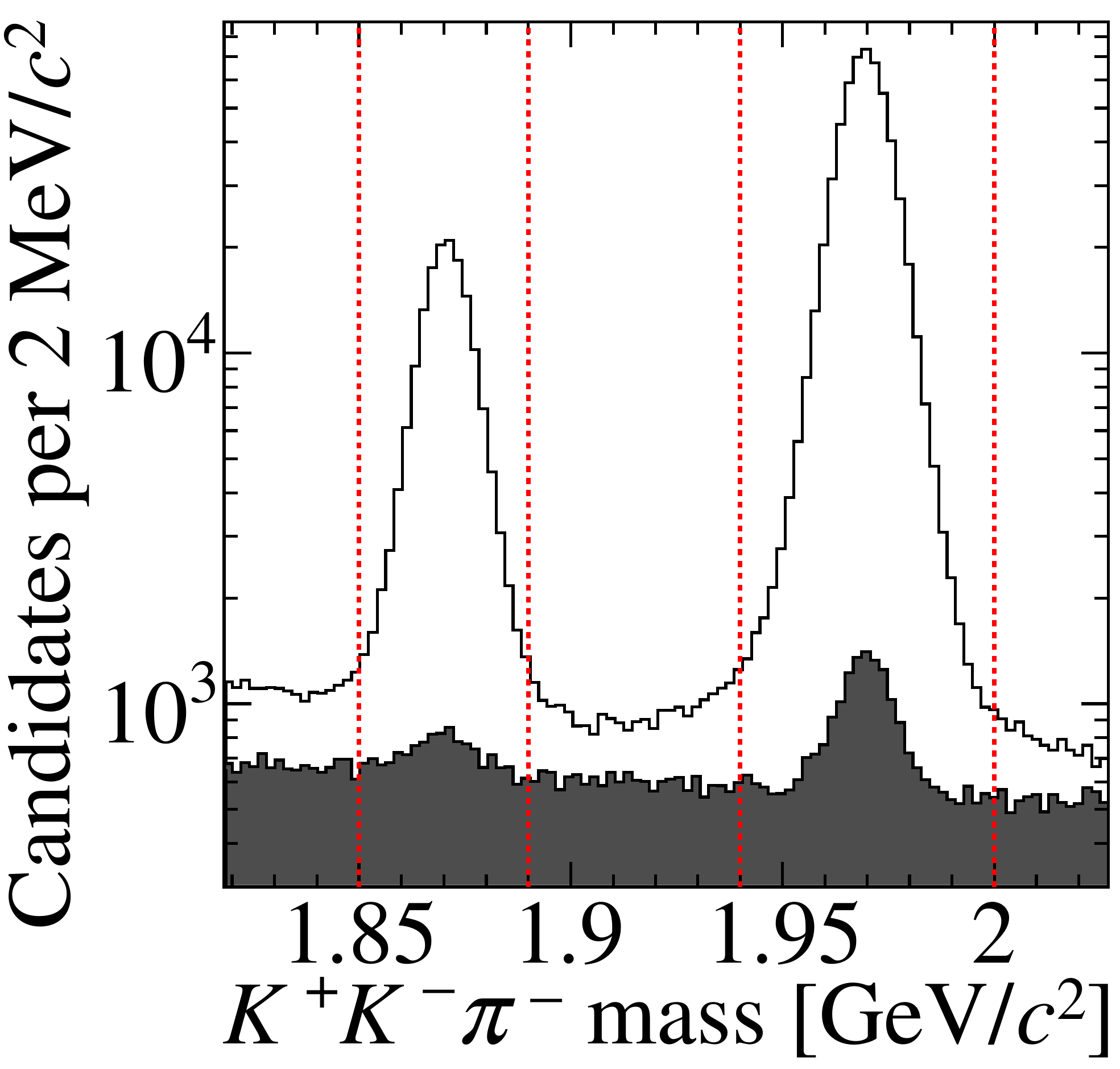}}
\end{overpic}\\
\caption{\label{fig:visibleMass} Distributions of $D\mu$ mass for (top panel) reference candidates, formed by combining $\Dm \to K^+K^-\pi^-$ candidates with $\mu^+$ candidates, and (bottom panel)  signal candidates formed by $\Dsm \to K^+K^-\pi^-$ candidates combined with $\mu^+$ candidates.  The inset shows the $K^+K^-\pi^-$-mass distribution with vertical lines enclosing the \Dm (\Dsm)  candidates used to form the reference (signal) candidates. The dark-filled histograms show same-sign candidate distributions.}
\end{figure}

Approximately 468\,000 (141\,000) signal (reference) candidates, formed by combining with candidates $\mu^+$ the $K^+K^-\pi^-$ candidates consistent with \Dsm (\Dm) decays, fulfill the selection. Figure~\ref{fig:visibleMass} shows the $D\mu$ mass distributions with corresponding $K^+K^-\pi^-$  mass distributions in the inset. \par In the $D\mu$ distribution, the enhancements of the signal and reference distributions over the corresponding same-sign distributions for $m_{\D\mu}< 5.5 \gevcc$ are predominantly due to bottom-hadron decays. The gap of candidates at $m_{\D\mu}\approx 5.3 \gevcc$ reflects the $B^0_{(s)} \to D^-_{(s)}\pi^+$ veto. The two peaks in the $K^+K^-\pi^-$ distributions of same-sign candidates are due to genuine charm decays accidentally combined with muon candidates. Along with $\Bs \to [K^+K^-\pi^-]_\DSSm \mu^+\nu_\mu$ decays, many \Bs decays potentially useful for the lifetime measurement contribute signal candidates, including decays into $D_{(s)}^{**}(\to \DSSm X) \mu^+ \nu_{\mu}$,  $\Dsm \tau^+ (\to \mu^+ \nu_{\mu} \bar{\nu}_\tau) \nu_\tau$, $\Dssm (\to \Dsm X) \tau^+ (\to \mu^+ \nu_\mu \bar{\nu}_{\tau}) \nu_{\tau}$, and $D^{**}_{s} (\to \DSSm X) \tau^+ (\to \mu^+\nu_\mu \bar{\nu}_{\tau}) \nu_\tau$ final states.\footnote{The symbol $D^{**}_{(s)}$ identifies collectively higher orbital excitations of $D^{-}_{(s)}$ mesons throughout.} Similarly, along with the $\Bz \to [K^+K^-\pi^-]_\DSm \mu^+\nu_\mu$ decays, potential reference candidates are contributed by \Bz decays into $D^{**}(\to D^{\scalebox{0.4}{(}*\scalebox{0.4}{)}-} X ) \mu^+ \nu_\mu$,  $D^- \tau^+ (\to \mu^+ \nu_\mu \bar{\nu}_\tau) \nu_\tau$, $D^{*-} (\to \Dm X) \tau^+ (\to \mu^+ \nu_\mu \bar{\nu}_\tau) \nu_\tau$, and $D^{**}(\to D^{\scalebox{0.4}{(}*\scalebox{0.4}{)}-} X )\tau^+ (\to \mu^+ \nu_\mu \bar{\nu}_\tau) \nu_\tau$ final states. However, to simplify the analysis we restrict the signal (reference) decays solely to the $\Bs \to [K^+K^-\pi^-]_\DSSm \mu^+\nu_\mu$  ($\Bz \to [K^+K^-\pi^-]_\DSm \mu^+\nu_\mu$) channels since they already contribute 95\% (91\%) of the inclusive $K^+K^-\pi^-\mu^+$ yield from semileptonic \Bz (\Bs) decays and require smaller and better-known $k$-factor corrections to relate the observed decay times to their true values.
\begin{figure}
\centering
\includegraphics[width=0.48\textwidth]{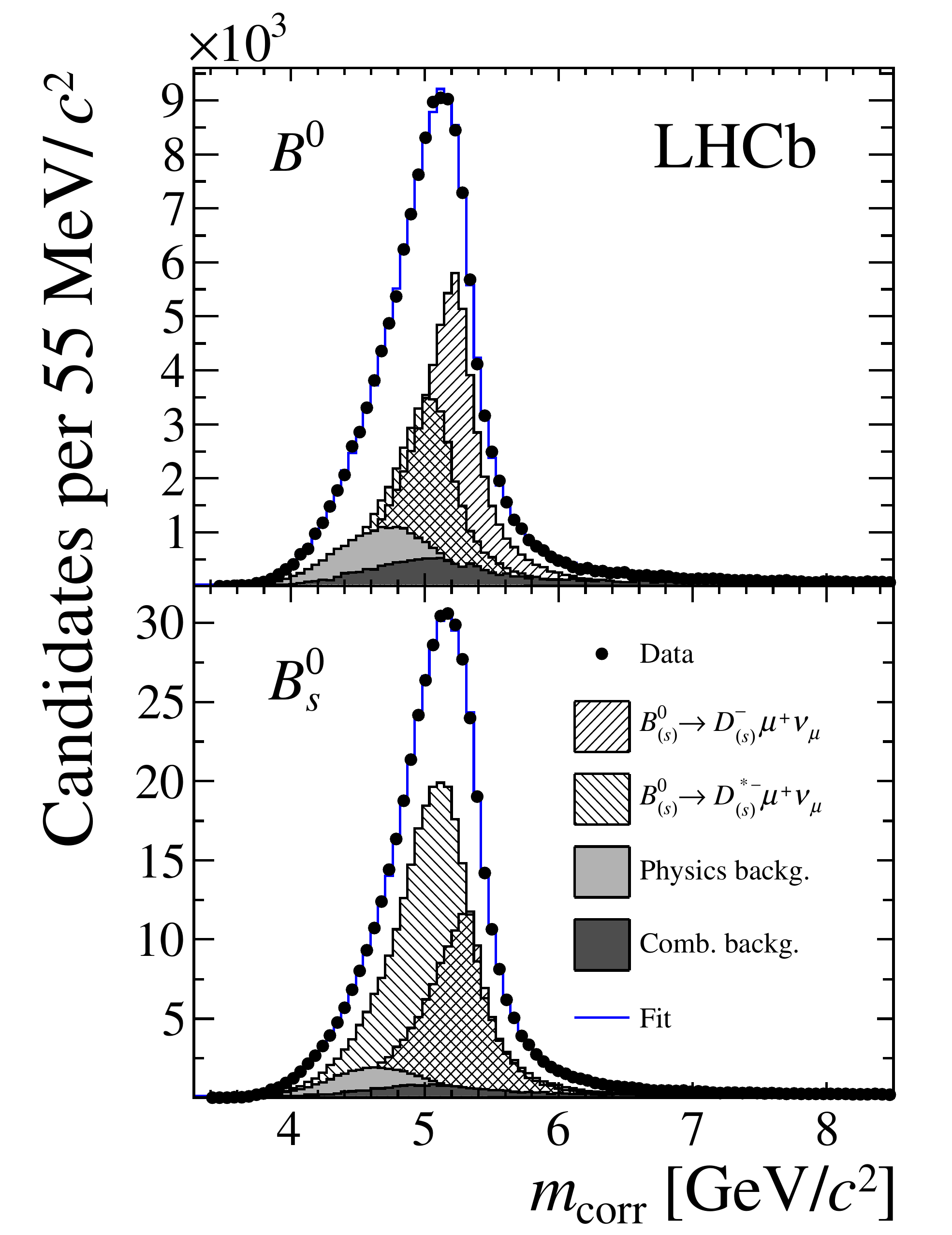}\\
\caption{Corrected-mass distributions for (top panel) reference $\Bz \to [K^+ K^- \pi^-]_{\DSm} \mu^+ \nu_\mu$ 
and (bottom panel) signal  $\Bs \to [K^+K^-\pi^-]_{\DSSm} \mu^+ \nu_\mu$ 
candidates satisfying the selection. Results of the global composition-fit are overlaid. In the \Bs fit projection, the lower- and higher-mass background components described in the text are displayed as a single, merged ``physics background" component. \label{fig:B_M_AfterSelection}} 
\end{figure}
\par A reliable understanding of the sample composition is essential for correct lifetime results.  An unbiased determination, from simulation, of the acceptances and mass distributions as functions of decay time requires that the composition of the simulated sample mirrors the data composition. We therefore weight the composition of the simulated samples according to the results of a global, least-squares composition fit to the $m_{\rm corr}$ distributions in data, shown in Fig.~\ref{fig:B_M_AfterSelection}. In the \Bs sample, such fit includes the two signal components, $\Bs \to [K^+K^-\pi^-]_\Dsm \mu^+ \nu_\mu$ and $\Bs \to [K^+K^-\pi^-]_\Dssm \mu^+ \nu_\mu$;  a combinatorial component; and two physics backgrounds. Each physics background component  is formed by grouping together processes yielding sufficiently similar corrected-mass distributions, resulting in a contribution at lower values of corrected mass ($\Bz \to D^{\scalebox{0.4}{(}*\scalebox{0.4}{)}-}D^{\scalebox{0.4}{(}*\scalebox{0.4}{)}+}_s$, $\Bp \to \Dbar{}^{\scalebox{0.4}{(}*\scalebox{0.4}{)}0}D^{\scalebox{0.4}{(}*\scalebox{0.4}{)}+}_s$, and $D^{**}(\to \DSSm X) \mu^+ \nu_\mu$) and another at higher corrected-mass values ($\Bp \to \DSSm K^+ \mu^+  \nu_\mu X$, $\Bz \to \DSSm \Kz \mu^+ \nu_\mu X$, and $\Bs \to \Dsm \tau^+ (\to \mu^+ \nu_\mu \bar{\nu}_\tau) \nu_\tau X$). All distributions are modeled empirically from simulation, except for the combinatorial distribution, which is modeled using same-sign data. Contributions expected to be smaller than 0.5\% are neglected. The impact of this approximation, and of possible variations of the relative proportions within each fit category, are accounted for in the systematic uncertainties. The fit has 62.1\% $p$-value and determines the fractions of each component with 0.13\%--0.91\% absolute statistical uncertainty. \par A simpler composition fit is used for the \Bz sample. Signal and combinatorial components mirror those of the \Bs case; the contributions from $\Bz \to D^{**-}(\to D^{\scalebox{0.4}{(}*\scalebox{0.4}{)}-} X )\mu^+ \nu_\mu$ and $\Bp \to \Dm \mu^+ \nu_\mu X$ decays have sufficiently similar distributions to be merged into a single physics-background component. The results of the corrected-mass composition fit of the reference sample, and of a sample of 2.1 million $\Bz \to [K^+\pi^-\pi^-]_\DSm \mu^+\nu_\mu$ decays where the  \Dm meson is reconstructed in the $K^+\pi^-\pi^-$ final state, offer a stringent validation. Discrepancies in the individual fractional contributions with respect to precise results from other experiments do not exceed 1.3 statistical standard deviations.\par
\begin{figure}[t]
\centering
\includegraphics[width=0.48\textwidth]{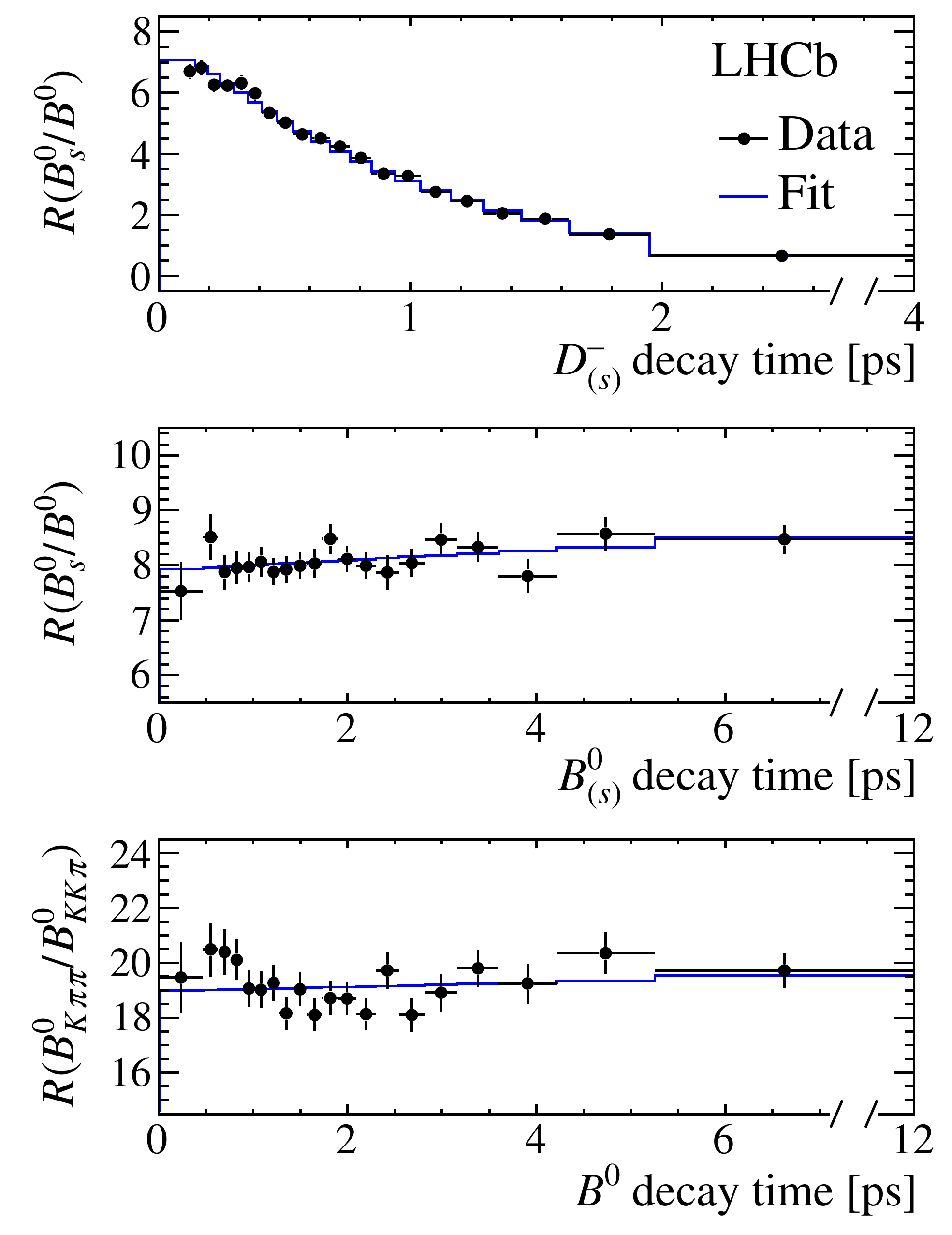}\\
\caption{Ratio between acceptance-corrected yields of signal  $\Bs \to [K^+K^-\pi^-]_{\DSSm} \mu^+\nu_\mu$ 
and reference $\Bz \to [K^+K^-\pi^-]_{\DSm} \mu^+\nu_\mu$  decay yields as a function of (top panel) charm-meson and (middle panel) bottom-meson decay time. The bottom panel shows the ratio between acceptance-corrected \Bz decay yields in the $[K^+\pi^-\pi^-]_{\DSm} \mu^+\nu_\mu$  and $[K^+K^-\pi^-]_{\DSm} \mu^+\nu_\mu$ channels as a function of \Bz decay time.  Fit results are overlaid. Relevant for the results is only the slope of the ratios as a function of decay time; absolute ratios, which depend on the decay yields, weighting, and efficiencies, are irrelevant. \label{fig:timefit_Bdratio}}
\end{figure}

The composition fit is sufficient for the determination of $\Delta_\Gamma(D)$, where no $k$-factor corrections are needed since the final state is fully reconstructed. We determine $\Delta_\Gamma(D)$ through a least-squares fit of the ratio of signal \Bs and reference \Bz yields as a function of the charm-meson decay time in the range 0.1--4.0\,ps. The yields of signal $\Bs \to [K^+K^-\pi^-]_\DSSm \mu^+\nu_\mu$ and reference $\Bz \to [K^+K^-\pi^-]_\DSm \mu^+\nu_\mu$ decays are determined in each of 20 decay-time bins with a $m_{\rm corr}$ fit similar to the global composition-fit. The two signal and the two physics-background contributions are each merged into a single component according to the proportions determined by the global fit and their decay-time evolution expected from simulation. The fit includes the decay-time resolution and the ratio between signal and reference decay-time acceptances, which are determined to be uniform within 1\%  from simulation.  The fit is shown in the top panel of Fig.~\ref{fig:timefit_Bdratio}; it has 34\% $p$-value and determines $\Delta_\Gamma(D) = 1.0131 \pm 0.0117$\invps.\par
\begin{figure}[t]
\centering
\includegraphics[width=0.48\textwidth]{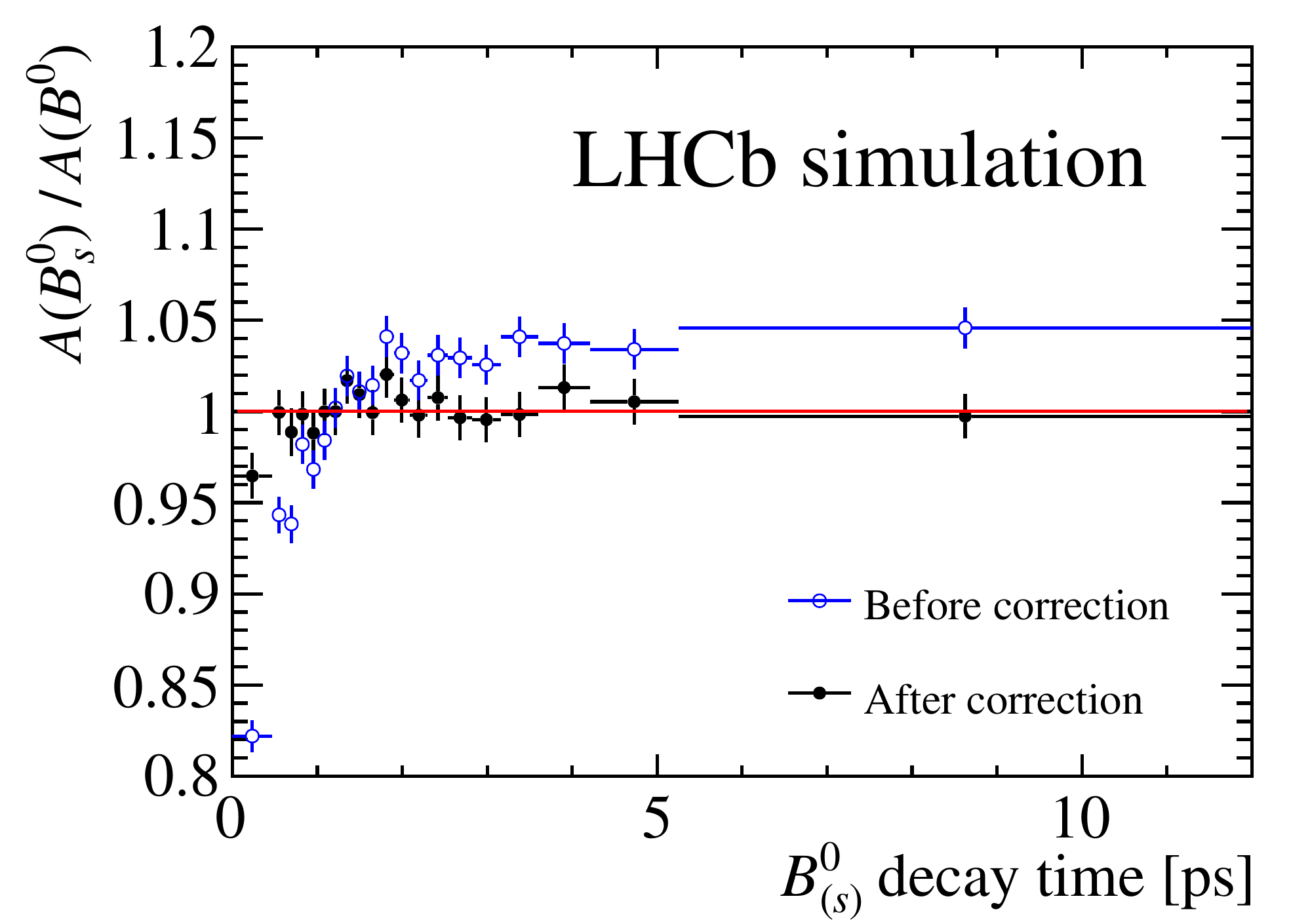}\\
\caption{Ratio between signal and reference decay acceptance as a function of decay time (open dots) prior to and (full dots) after the acceptance correction, with the result of a fit of the latter overlaid.\label{fig:acceptance}}
\end{figure}

\begin{figure}[t]
\centering
\includegraphics[width=0.48\textwidth]{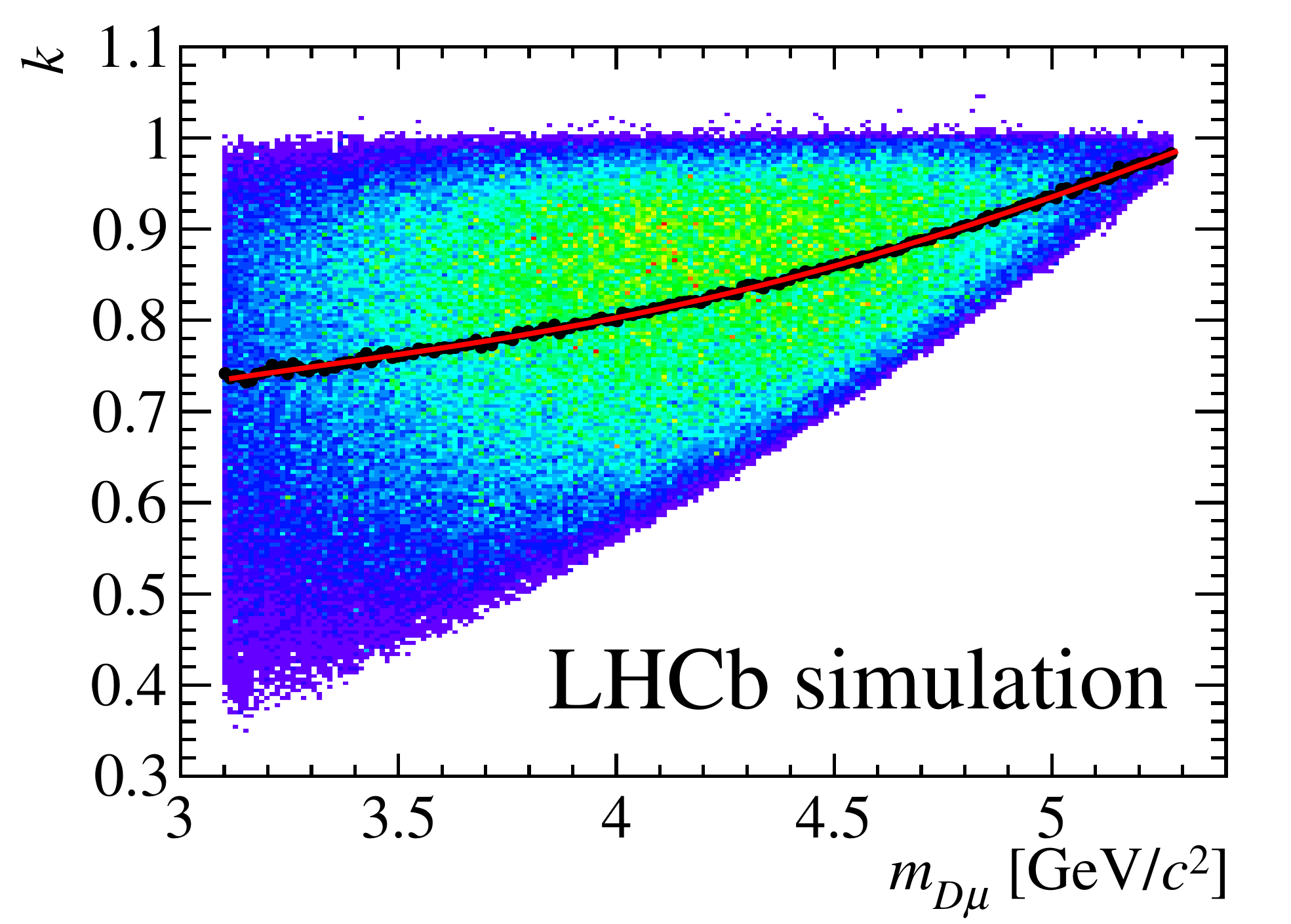}\\
\caption{Distribution of $k$-factor as a function of $m_{D\mu}$ in simulated signal data with an empirical fit of its $m_{D\mu}$-averaged value overlaid. \label{fig:kfactor}}
\end{figure}

 
The measurement of $\Delta_\Gamma(B)$ requires, in addition, an acceptance correction for the differences between signal and reference decays, and the $k$-factor correction. The acceptance correction accounts for the difference in decay-time-dependent efficiency due to the combined effect of the difference between \Dm and \Dsm lifetimes and the online requirements on the spatial separation between $D^-_{(s)}$ and $B^0_{(s)}$ decay vertices: we apply to the \Bs sample a per-candidate weight, $w_i \equiv \exp[\Delta_\Gamma(D) t(\Dsm)]$, based on the $\Delta_\Gamma(D)$ result and the \Dsm decay time, such that the \Dsm and \Dm decay-time distributions become consistent. Figure~\ref{fig:acceptance} shows the effect of the weighting on the acceptance.
The $k$-factor is the average fractional contribution of the observed momentum to the true momentum determined in a simulated sample. The $k$-factor-dependence on the kinematic properties of each candidate is included through a dependence on $m_{D\mu}$,  $k(m_{D\mu}) = \left\langle p_{D\mu}/p_{\rm true}\right\rangle$, where $p_{\rm true}$ indicates the true momentum of the $B^0_{(s)}$ meson (Fig.~\ref{fig:kfactor}).
Our candidate-specific correction consists in dividing the candidate's momentum reconstructed in data by the $k$-factor.  
Equalized compositions of simulated and experimental data samples ensure that the $k$-factor distribution specific to each of the four signal and reference decays is unbiased. \par We determine $\Delta_\Gamma(B)$ with the same fit of $m_{\rm corr}$ used to measure $\Delta_\Gamma(D)$ except that here the ratios of signal and reference yields are determined as functions of the $B^0_{(s)}$ decay time. The decay-time smearing due to the $k$-factor spread is included in the fit.  After the \Dsm lifetime weighting, the decay-time acceptances of simulated signal and reference modes are consistent, with a $p$-value of $83\%$, and are not included in the fit. The fit is shown in the middle panel of Fig.~\ref{fig:timefit_Bdratio}; the resulting width difference is $\Delta_\Gamma(B) =  -0.0115 \pm 0.0053$\invps, with 91\% $p$-value.

We validate the analysis with a null test to check against biases due to differences in acceptances and kinematic properties,  We repeat the width-difference determination by using the same reference $\Bz \to [K^+K^-\pi^-]_\DSm \mu^+\nu_\mu$ sample and replacing the signal decays with $\Bz \to [K^+\pi^-\pi^-]_\DSm \mu^+\nu_\mu$ decays, where the \Dm meson is reconstructed in the $K^+\pi^-\pi^-$ final state.   Differing momentum and vertex-displacement selection criteria induce up to 10\% acceptance differences as a function of $\Dm$ decay time and up to 25\% variations as a function of \Bz decay time. Acceptance ratios are therefore included in the fit (Fig.~\ref{fig:timefit_Bdratio}, bottom panel). The $p$-values are 21\% for the \Bz fit and 33\% for the \Dm fit. The resulting width differences, $\Delta_\Gamma(D) = (-19 \pm 10)\times 10^{-3}$\invps and $\Delta_\Gamma(B) = (-4.1 \pm 5.4)\times 10^{-3}$\invps, are consistent with zero, hence supporting the overall validity of the approach.

We assess independent systematic uncertainties due to (i) potential fit biases; (ii) assumptions on the components contributing to the sample and their mass distributions; (iii) assumptions on the signal decay model, e.g., choice of $\Bs \to \Dssm$ form factors; (iv) uncertainties on the decay-time acceptances; (v) uncertainties on the decay-time resolution; (vi) contamination from \Bs candidates produced in $B_c^+$ decays; and (vii) mismodeling of the expected \pt differences between \Bz and \Bs mesons.  We evaluate each contribution by including the relevant effect in the model and repeating the whole analysis on ensembles of simulated experiments that mirror the data. For the $\Delta_\Gamma(D)$ result, the systematic uncertainty is dominated by a 0.0049\invps contribution due to the decay-time acceptance, and a 0.0039\invps contribution due to the decay-time resolution. A smaller contribution of 0.0018\invps arises from possible mismodeling of \pt differences in \Bz and \Bs production. For the $\Delta_\Gamma(B)$ result,  a 0.0028\invps uncertainty from mismodeling of \pt differences between \Bz and \Bs mesons and a 0.0025\invps contribution from the \Bs decay model dominate. Smaller contributions arise from \Bc feed-down (0.0010\invps), residual fit biases (0.0009\invps), sample composition (0.0005\invps), and decay-time acceptance and resolution (0.0004\invps each). The uncertainties associated with the limited size of simulated samples are included in the fit $\chi^2$ and contribute up to 20\% of the statistical uncertainties. The uncertainty in the decay-length measurement has negligible impact. Consistency checks based on repeating the measurement independently on subsamples chosen according to data-taking time, online-selection criteria, charged-particle and vertex multiplicities, momentum of the $K^+K^-\pi^-\mu^+$ system, and whether only the $\Dsm \mu^+ \nu_\mu$ or the $\Dssm \mu^+ \nu_\mu$ channel is considered as signal, all yield results compatible with statistical fluctuations.
\section{Summary of results and discussion}
We report world-leading measurements of \Bs and \Dsm meson lifetimes using a novel method. We reconstruct $\Bs \to \Dssm \mu^+\nu_\mu$ and $\Bs \to \Dsm \mu^+\nu_\mu$ decays in proton-proton collisions collected by the LHCb experiment and corresponding to 3.0\invfb of integrated luminosity. We use $\Bz \to \Dstarm \mu^+\nu_\mu$ and $\Bz \to \Dm \mu^+\nu_\mu$ decays reconstructed in the same final state as a reference to suppress systematic uncertainties. The resulting width differences are $\Delta_\Gamma(B) = \dB\pm \eStatdB\stat \pm \eSystdB\syst$\invps and $\Delta_\Gamma(D) = \dD\pm \eStatdD\stat \pm \eSystdD\syst$\invps. They are uncorrelated. Using the known values of the \Bz~\cite{PDG2016, Aaij:2014owa} and \Dm lifetimes~\cite{PDG2016,Link:2002bx}, we determine the flavor-specific \Bs lifetime, $\taufs = \tB \pm \eStattB\stat \pm \eSysttB\syst\pm\eReftB\refB$\,ps, and the \Dsm lifetime, $\tau_{\Dsm} = \tD \pm \eStattD\stat \pm \eSysttD\syst\pm \eReftD\refD$\,ps; the uncertainties are dominated by the size of the reference sample, and the last contributions are due to the uncertainties on the \Bz and \Dm lifetimes, respectively. \par The results improve by 15\% over the current \taufs\ value and by a factor of two the current $D^+_s$ lifetime, whose precision had not been improved in the past decade~\cite{Link:2005ew,Abazov:2014rua,LHCb-PAPER-2014-037}. They might offer improved insight into the interplay between strong and weak interactions in the dynamics of heavy mesons and sharpen the reach of indirect searches for non-standard-model physics. \par Promising opportunities of improvement are available. Extensions to events collected by additional triggers may offer an approximate 20\% increase in signal yield from the same data used in this work; addition of the 2015--2019 LHCb data set will further triple the signal yields; usage of higher-yield reference decays, like $\Bz \to [K^+\pi^-\pi^-]_\DSm \mu^+\nu_\mu$, will further reduce the statistical uncertainty. This work enables again, after a decade of declining interest, the opportunity of using semileptonic decays to achieve competitive measurements of lifetimes and other observables, like semileptonic branching fractions or \Bs\ form factors,  in LHCb and other experiments.

%
%

\bigskip
I thank Andreas Kronfeld, Alexander Lenz, Jonathan Rosner, and G.~Punzi for useful discussions.

%
%
%
%
%
%

\addcontentsline{toc}{section}{References}
\setboolean{inbibliography}{true}
\ifx\mcitethebibliography\mciteundefinedmacro
\PackageError{LHCb.bst}{mciteplus.sty has not been loaded}
{This bibstyle requires the use of the mciteplus package.}\fi
\providecommand{\href}[2]{#2}


\def\Discussion{
\setlength{\parskip}{0.3cm}\setlength{\parindent}{0.0cm}
     \bigskip\bigskip      {\Large {\bf Discussion}} \bigskip}
\def\speaker#1{{\bf #1:}\ }
\def\endDiscussion{}

%
%
%
%
 
\end{document}